\documentclass[review]{elsarticle}

\usepackage{lineno,hyperref}
\usepackage{amssymb}
\usepackage{amsmath}
\usepackage{algorithmic}
\usepackage{algorithm}
\usepackage{textcomp}
\usepackage{listings,xcolor}
\usepackage{soul}
\usepackage{tabularx}
\usepackage{setspace}

\newcommand{\changed}[1]{\noindent\textcolor{blue}{ #1}}

\journal{Journal of Biomedical Informatics} 

\begin{document}

\begin{frontmatter}

\title{Predicting future state for adaptive clinical pathway management}

\author[agfa]{Hong Sun}
\author[iminds,dresden]{D\"orthe Arndt}
\author[agfa]{Jos De Roo}
\author[iminds]{Erik Mannens}

\address[agfa]{
Agfa Healthcare --
Moutstraat 100,
9000 Ghent, Belgium
}

\address[iminds]{%
  IDLab, Department of Electronics and Information Systems, Ghent University -- imec, 
AA-Tower, Technologiepark 122, B-9052 Ghent, Belgium
}
\address[dresden]{%
  Computational Logic Group, TU Dresden, Germany
}

\begin{abstract}
 %%\da{question: does the journal have any limitations on the length of the abstract?}
 %%\hsun{It didn't mention any limitation on abstract length, are you thinking it is tool long or too short? It is currently 262 words, I think it is appropriate, and I consider the last sentence can be skipped if you consider it as too long}
Clinical decision support systems are assisting physicians in providing care to patients. However, in the context of clinical pathway management such systems are rather limited as they only take the current state of the patient into account and ignore the possible evolvement of that state in the future. In the past decade, the availability of big data in the healthcare domain did open a new era for clinical decision support. Machine learning technologies are now widely used in the clinical domain, nevertheless, mostly as a tool for disease prediction. A tool that not only predicts future states, but also enables adaptive clinical pathway management based on these predictions is still in need. This paper introduces weighted state transition logic, a logic to model state changes based on actions planned in clinical pathways. Weighted state transition logic extends linear logic by taking weights --- numerical values indicating the quality of an action or an entire clinical pathway --- into account. It allows us to predict the future states of a patient and it enables adaptive clinical pathway management based on these predictions. We provide an implementation of weighted state transition logic using semantic web technologies, which makes it easy to integrate semantic data and rules as background knowledge. Executed by a semantic reasoner, it is possible to generate a clinical pathway towards a target state, as well as to detect potential conflicts in the future when multiple pathways are coexisting. The transitions from the current state to the predicted future state are traceable, which builds trust from human users on the generated pathway.

\end{abstract}


\end{frontmatter}

%%\linenumbers

\section{Introduction}

Clinical pathways are tools used to guide evidence-based healthcare to promote organised and efficient patient care \cite{kinsman2010clinical,de2006defining, rotter2010clinical}. They follow disease-specific guidelines to coordinate a set of services to be executed by different stakeholders, and aim to optimise outcomes in settings such as acute care and home care \cite{deneckere2012care}.  A clinical pathway provides a set of treatments/actions that help a patient to move from the current state to a target state that a disease is cured, or controlled. 

Clinical pathways are widely used in hospitals to translate clinical practice guidelines into  clinical processes of care within the unique environment of a healthcare institution \cite{rotter2019clinical}. It was reported in 2003 that more than 80\% of hospitals in the USA had implemented clinical pathways \cite{rotter2010clinical}. In 2019, it was reported that clinical pathways were used in most European countries, and there were increasing activities in the development and implementation of clinical pathways in some countries such as Belgium and Germany \cite{rotter2019clinical}.

Today, clinical pathways are typically defined based on best practices by multidisciplinary teams within one care organization, for a specific disease, and for a typical patient profile \cite{chevalley2002osteoporosis,takegami2003impact}. The pathways are finally shared on paper or integrated into operational clinical IT systems and impose the actions to be performed by caregivers and patients. Such pathways cannot span over multiple organizations and are often considered too generic and inflexible for adaptation to the characteristics or situations of an individual patient.

The care plan made by a traditional clinical pathway management tool is mostly built with the current state and a disease to be tackled. It lacks a global vision of the possible future state, which is a combined consequence of several pathways. In a situation like comorbidity, there coexist several clinical pathways to cope with different diseases. A planned action in one pathway would change the state of a patient in the future, and such a change is not explicit to another pathway, which is only produced based on the current state. In some situations, such a change of patient state might trigger a risk or alarm for another pathway. While some of such conflicts can be detected by stating certain actions as conflicts, e.g., contraindication, others are not obvious at the planning stage, but only manifest themselves when the action is carried out with visible consequences. One example of such a conflict is that a patient is scheduled for an X-ray without contrast fluid on one day for a check-up of disease A, and later is scheduled with an extra X-ray with contrast medium one day earlier for a check-up of disease B. In such a case, the X-ray for disease A cannot be taken anymore because of the contamination by the contrast medium and will need to be postponed until the contrast is washed out from the body. If a clinical decision support tool is able to foresee expected future states, such a conflict would be detected before it actually takes place. 

This paper introduces weighted state transition logic and an implementation of that logic based on semantic web technology. It allows us to explicitly describe the expected state transitions of a planned clinical event in N3 logic \cite{berners2008n3logic}. Executed by the semantic reasoning engine EYE \cite{eye}, it is possible to generate clinical pathways towards a target state. In addition, it is able to predict potential conflicts in the future when multiple pathways are coexisting, without needing to actually execute those actions. Moreover, by assigning weights to each step, it is possible to generate overall weights for each possible pathway from the current state to a target state, 
%so as to assist path selection
and thereby to support informed path selection.

The presented logic was developed and implemented in the GPS4IC (GPS for Integrated Care) project \cite{GPS4IC} to generate adaptive personalized clinical pathways. By introducing services from home care providers, the system extends its scope from hospital to ambient assisted home environments \cite{sun2009promises} to provide integrated care \cite{campbell1998integrated}. 

The remainder of this paper discusses the concept of weighted state transition logic, followed by an implementation of that logic using semantic web technology. We explain the architecture of our implementation as it was applied in the GPS4IC project, and provide the link to an example scenario on GitHub. Lessons learned regarding the application of weighted state transition logic are given at the end of this paper.

\section{Related work}

The target of developing tools supporting adaptive clinical pathways to cope with comorbidity and to provide personalized care has attracted several research initiatives. They inspired the creation of the weighted state transition logic presented in this paper.

Predesigned clinical pathways are often focused on one specific disease and lack the support for the treatments of comorbidities and complications \cite{campbell1998integrated, rotter2010clinical, HUANG2016227}. Huang et al. applied latent pattern mining to detect comorbidities in clinical pathways, however, they did not provide solutions to adapt the clinical pathways accordingly \cite{HUANG2016227}. Colaert et al. \cite{colaert2007,chen2004towards} introduced the term adaptive clinical pathway, which goes beyond classical clinical pathway. It is about an intelligent and federated workflow system using semantic technology, crossing the episodic and local hospital boundaries to come to a life long and regional healthcare system.  

Sun et al. \cite{sun2015semantic} built a virtual semantic layer on top of Electronic Health Records (EHRs), to integrate healthcare data and to support different clinical research. Zhang et al. \cite{zhang2016integrating} proposed a unified representation of healthcare domain knowledge and patient data based on HL7 RIM and ontologies, and developed a semantic healthcare knowledge base. Both works built the foundation to represent and process healthcare data in a common way that allows the application of semantic rules by a reasoning engine.

Alexandrou et al. \cite{alexandrou2010holistic} implemented the SEMPATH software platform, which leverages the provision of highly personalized health care treatment by utilizing and managing clinical pathways. SEMPATH performs rule-based exception detection with the semantic web rule language (SWRL) \cite{swrl}. It performs dynamic clinical pathway adaptation during the execution time of each pathway to personalize the treatment scheme. Wang et al. \cite{wang2013creating} semantically processed and aggregated EHRs with ontologies relevant for clinical pathways. They applied reasoning by rules in SWRL to adjust standardised clinical pathways to meet different patients' practical needs. Both works can cope with comorbidity and provide personalized care with semantic rules. The limitation is that the event-driven adaption is only triggered when the patient state changes. It is not feasible in advance to predict and avoid a potential conflict that may pop up in the future.

Using machine learning (ML) technology to construct predictive modelling with EHRs is attracting much research interest in recent years. Rajkomar et al. \cite{rajkomar2018scalable} build scalable and accurate deep learning with EHRs to predict a set of clinical events. The limitation of applying ML technology in adaptive clinical pathway management is that ML is mostly focused on predicting an upcoming clinical event, and lacking the ability to make use of the predicted event for adaptive clinical pathway management. In addition, issues such as explainability and transparency of the machine learning models still need to be addressed \cite{cutillo2020machine}. 

Bradbrook et al. \cite{bradbrook2005ai} investigated applying AI planning technology in clinical practice guidelines, and suggested that techniques such as Planning Domain Definition Language (PDDL) \cite{mcdermott1998pddl} and PROforma \cite{sutton2003syntax} may make a substantial contribution to computerised care plan representation and execution. Alaboud et al. \cite{alaboud2019personalized} applied PDDL+ \cite{fox2002pddl+} in planning the daily routine of a patient in terms of pain relief medications and activities such as eat and drive. These research findings are important efforts in describing state transitions of the care-path management. The limitation is that those technologies are still not sufficient to describe the complex states in care-path management, or to cope with situations such as interactions between medications and constraints on medication usage. 
 
Verborgh et al. proposed a method (RESTdesc) \cite{verborgh2017pragmatic, verborgh2014serendipitous} to automatically find a path to reach a specific goal by executing a set of steps sequentially. RESTdesc is built with semantic web technology and describes the states as well state transitions in N3 logic \cite{berners2008n3logic}. It inherits the expressive and reasoning power, as well as the openness in linking data. By stating the current states as facts, the steps as rules, and the target state as a query, a semantic reasoning engine is able to find a satisfactory path to lead from the current state to the target state. There are still limitations that prevent the application of RESTdesc in the clinical domain, which will be discussed in detail in the next section. Weighted state transition logic as presented in this paper is inspired by RESTdesc, and made several significant improvements to fit the requirements of clinical pathway generation and adaption. 

\section{Modeling state transition in clinical domain}

\subsection{The requirements}
A clinical pathway is one of the main tools to manage the delivery of quality care. It is often following standardized guidelines and it consists of a sequence of treatments/actions that helps a patient to move from the current state to a target state that a disease is cured, or controlled. Although the starting state is explicit as being the current state, the intermediate states are implicit because each treatment/action would change the state of the patient who is receiving it. Even if a treatment is only meant to keep the current state, it leads to a new state that such a treatment is received. While a physician is designing a clinical pathway, the consequence of each action is implicitly applied in the mind of the physician. However, when dealing with comorbidity, where there are multiple pathways that are dealing with different diseases, it is difficult and time consuming for a physician to take into account the consequences of each action that is planned by other physicians. Although some clinical decision support tools are able to detect drug contraindications, it is extremely difficult to detect conflicting events that are scheduled to be carried out in the future. Enabling clinical decision support tools to make an analysis not only based on the current facts, but also taking into account the influences of the existing pathways (i.e. planned actions) becomes a requirement, as well as a challenge.

In order to allow clinical decision support tools to take into account the influences of the existing pathways, it is a prerequisite to make such influences explicit. While the sequences to execute the actions are usually explicitly stated, the consequences of these executions are often implicit. The consequence of an action represents the expected future state, and the expected future state can be propagated when more than one action is planned sequentially. By taking into account the consequences of planned actions, the weighted state transition logic presented in this paper is able to predict the future state that allows us to perform tasks such as path generation and path validation. During the process of path generation and path validation, the model continuously updates its present state with the predicted future state. The process has the Markov property \cite{gurvits2005markov}: the conditional probability distribution of future states of the process depends only upon the present state, not on the sequence of events that preceded it. In path validation, the path to be validated is already given. At each step, the next action to execute is explicitly stated in the given path. In path generation, the path generation engine may choose any action that is executable at each step to move towards a new state. Such a process repeats until the target state is reached, and it is considered a Markov decision process \cite{howard1960dynamic}.

RESTdesc, as introduced in the related work, is also able to explicitly describe the consequence of an action as well as to generate a path towards a target state. Yet it still has two major limitations. Firstly, the state transition logic of RESTdesc only allows to assert new states and lacks support to retract old states. In the application of the clinical domain, it requires a state transition logic to retract statements that are no longer valid, e.g., the temperature or a lab test of a patient. Secondly, the RESTdesc solution only generates one path that leads to the target state, it does not display alternative paths. In the clinical decision support system, it is a preferred feature to provide alternative paths for path selection, ideally with overall weights (e.g., regarding cost, treatment time, etc.) of each path explicitly stated. 

The weighted state transition logic presented in this paper meets the aforementioned challenges. We first introduce some existing logic to model state change, then followed by our weighted state transition logic.

\subsection{Existing logic to model state change}

There are different ways to model state change. In this section, we analyze the limitations of some existing logic, and propose our weighted state transition logic to model state change.

\subsubsection{Classical and intuitionistic logic}

Classical and intuitionistic logic \cite{van1986intuitionistic} generates new states as follows:\\

%\newline

%	\[
%	  If A and A => B, then B, but A still holds
%	\]

\fbox{%
  \parbox{0.9\textwidth}{%
  If A and A $\Rightarrow$ B, then B, but A still holds
}%
}
\newline

In the intuitionistic implication, when the premise of an inference is fulfilled, the conclusion is derived, and the premise still holds as stable truth. This is correct in mathematics, but could be problematic in real-life, e.g., clinical applications, for example:

\begin{itemize}
\item  Let A  be the fact that patient X has temperature 40 \textdegree{}C
\item  Let B  be the fact that patient X has temperature 37 \textdegree{}C
\item Let $A \Rightarrow B$ , be the action of taking pill Paracetamol, that is the temperature drops from 40 \textdegree{}C (A) to 37 \textdegree{}C (B).
\end{itemize}

Then given A, following the intuitionistic implication  $A\changed{\Rightarrow}B$ (i.e. patient X has temperature 40 \textdegree{}C and takes pill Paracetamol), the consequence would be both fact A and fact B. The patient is with temperature both 40  \textdegree{}C and 37 \textdegree{}C, while in reality, only the latter is required and the former one is not needed any more.

\subsubsection{Linear logic}
Linear Logic solves this problem by eliminating the previous state. In Linear Logic \cite{girard1987linear, girard1995linear}, the state change is expressed as below, the fact on the left side of the transition will be consumed, and does not hold any more:
\newline

\fbox{%
  \parbox{0.9\textwidth}{%
      If A and A $\multimap B$, then B, and A does not hold any more 
}%
}
\newline

Take the aforementioned example:

\begin{itemize}
\item  Let A  be the fact that patient X has temperature 40 \textdegree{}C
\item  Let B  be the fact that patient X has temperature 37 \textdegree{}C
\item Let $A \multimap B$ be the action of taking pill Paracetamol, that is the temperature drops from 40 \textdegree{}C (A) to 37 \textdegree{}C (B).
\end{itemize}

Then given A and $A \multimap B$ (i.e. patient X has temperature 40 \textdegree{}C and takes pill Paracetamol), following linear logic, the consequence would be only fact B. The patient is with temperature 37 \textdegree{}C, and the fact that the patient is with temperature 40 \textdegree{}C will be dropped. 

However, if we modify the condition as follows:
\begin{itemize}
\item Let A  be the fact that patient X has temperature 40 \textdegree{}C, and patient X has no contraindication with Paracetamol.
\item  Let B  be the fact that patient X has temperature 37 \textdegree{}C
\item Let $A \multimap B$ be the action of taking pill Paracetamol, that is the temperature drops from 40 \textdegree{}C (A) to 37 \textdegree{}C (B).
\end{itemize}

Then given A and $A \multimap B$ (i.e. patient X has temperature 40 \textdegree{}C and takes pill Paracetamol),  following linear logic, the consequence would be fact B. The patient is with temperature 37 \textdegree{}C. The fact that the patient is with temperature 40 \textdegree{}C is dropped. However, the fact that the patient has no contraindication with Paracetamol will be dropped as well. This is against the original purpose, and missing this fact would prevent the drug Paracetamol being able to be applied in the future. 

To cope with such situations, linear logic also allows to express stable truth as the intuitionistic implication. $(!A) \multimap B$ is equivalent to $A \Rightarrow B$, it is therefore possible to introduce stable truth in linear logic with the expression below:\\

%\newline
\fbox{%
  \parbox{0.9\textwidth}{%
      If A and $(!A) \multimap B$, then B, and A still holds
}%
}
\newline

\subsection{Weighted state transition logic}
The state transition logic presented in this paper is largely inspired by linear logic. It uses the linear implication to express the state change, and also relies on intuitionistic implication to indicate stable truths. Below, we first give an informal introduction to our theory to clarify the idea behind. After that we complete our introduction by providing the corresponding definitions.

\subsubsection{Informal introduction}

\begin{figure*}[ht]
\centering\includegraphics[width=0.75\linewidth]{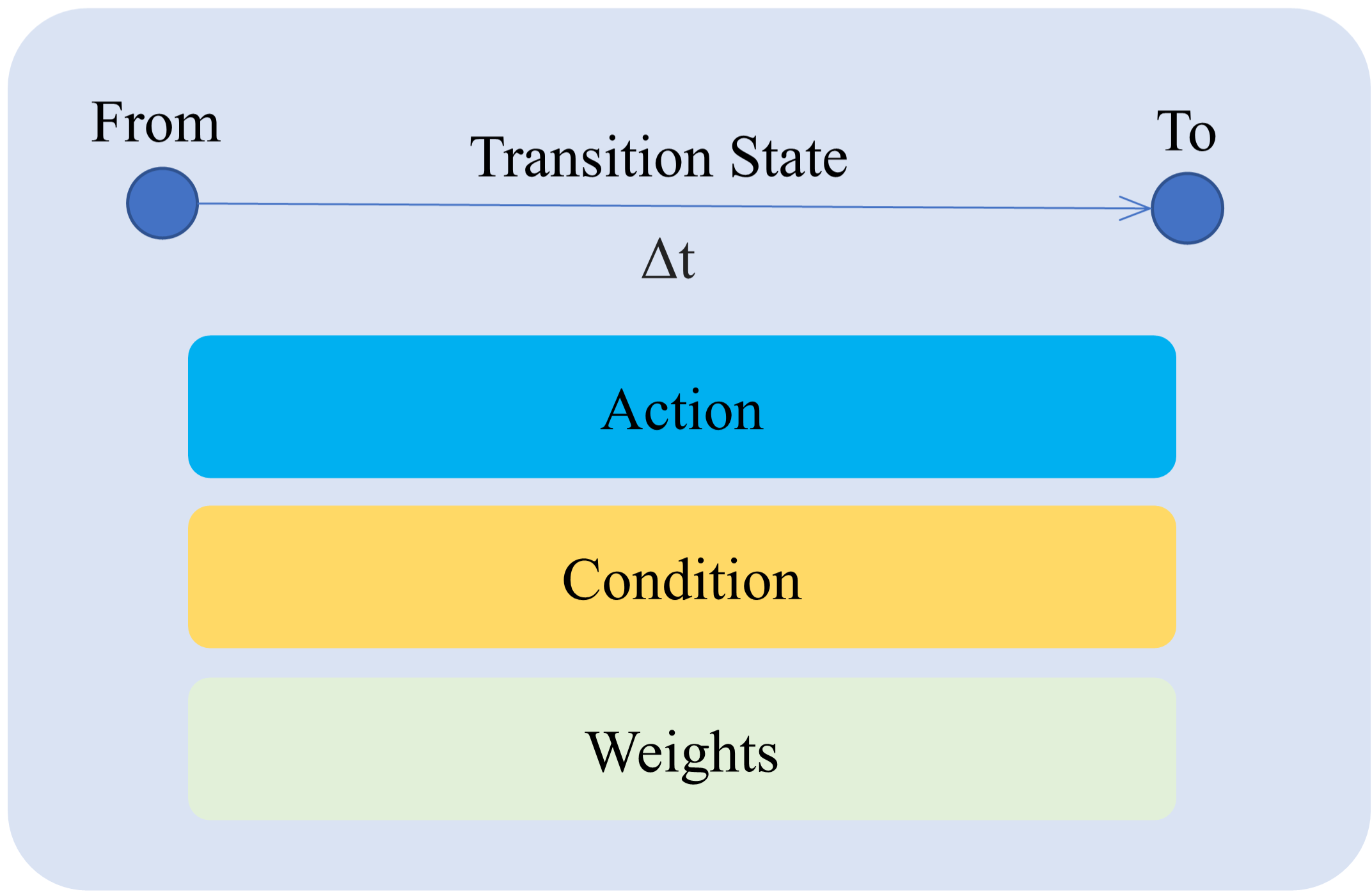}
\caption{Weighted state transition logic}
\label{fig:state-transition}
\end{figure*}

We start our considerations with a schematic overview. Figure \ref{fig:state-transition} shows the concept of weighted state transition logic. 'From' is representing the current state that is to be retracted, and 'To' is representing target state to be asserted. The section of the condition is representing the prerequisite to be fulfilled to carry out the state transition. The 'Condition' will not be retracted during the state transition. Below is the simplified representation of the proposed weighted state transition in terms of linear logic:\\

%\newline
\fbox{%
  \parbox{0.9\textwidth}{%
%      $ From \land (!Condition) \multimap To$
      $\text{From} \otimes (!\text{Condition}) \multimap \text{To}$
}%
}
\newline

The previous example of lowering body temperature can be expressed with the weighted state transition logic as follows:
\begin{itemize}
\item Let 'From' be the fact that patient X has temperature 40 \textdegree{}C 
\item Let 'To' be the fact that patient X has temperature 37 \textdegree{}C
\item Let 'Condition' be the fact that patient X has no contraindication with Paracetamol.
\item Let $\text{From} \otimes (!\text{Condition}) \multimap \text{To}$ be the fact that with the condition of patient X has no contraindication with Paracetamol, the action of taking pill Paracetamol has consequence that the temperature drops from 40 \textdegree{}C (A) to 37 \textdegree{}C (B).
\end{itemize}

It is important to emphasize that the condition section, patient X has no contraindication with Paracetamol, remains true after the state change. Moreover, it is also possible to add more complex expression of constraints or calculate the predicted values in the target state in the condition section. For example, the target temperature could be expressed in a more generic way as 3 \textdegree{}C less compared with the initial state, and the new value of '3 \textdegree{}C less' is calculated in the condition section. It could also state that this is only true when patient x has temperature more than 39.5 \textdegree{}C. An example of calculating target values is given in Listing 1 of Section \ref{semantic_description}. We deem the ability of proposed weighted state transition logic in expressing complex constraints and target state calculations as the advantage over the traditional linear logic.

It is also important to point out that in our weighted state transition logic, every fact only occurs once, while linear logic allows multiple occurrences of one fact. In addition, the weighted state transition logic presented in this paper allows to generate personalized adaptive clinical workflows, with the features listed in the subsections below.

\subsubsection{Duration of state change}
Temporal constraint management of clinical events is a crucial task in clinical pathway management \cite{combi2014representing}. It is important to know the start time as well as the duration of each step listed in a clinical pathway. Weighted state transition logic introduces the concept of duration to indicate the required time to complete the state transition, as is denoted by $\Delta$ T in Figure \ref{fig:state-transition}. The use of a duration allows to explicitly indicate the start and end time of a state change after the start time for the whole path is set up. It is also important to clarify that the duration is referring to the period of completing the state change, not the period of completing the action. For example, if Paracetamol is taken to lower the body temperature, the action of taking a pill is a point in time, but it takes two hours to lower the temperature by 2-3 degree. The duration of such a state change is then considered as two hours.

\subsubsection{Transition state of state change}
The weighted state transition logic is created to define and manage the change of states. The 'From' state is retracted, and the 'To' state is asserted. With the introduction of duration, the state change is no longer considered as an instantaneous event, but a transition which has a duration. When it takes two hours to lower the body temperature as described in the example case, it becomes unclear on when to retract the old state and when to assert the new state, as well as the state during the change. The concept of the transition state is introduced to make an explicit description of the state during the state transition period. At the start of the state change, the 'From' state is retracted, and the transition state is asserted. At the end state, the transition state is retracted, and the 'To' state is asserted. For example, once a pill of Paracetamol is taken, the 'From' state, that the patient is with 40 \textdegree{}C is retracted, and the transition state, that the patient has taken a pill of Paracetamol is asserted. After a duration of two hours, the transition state is retracted, and the 'To' state, that the patient is with 37 \textdegree{}C is asserted.

\subsubsection{Weights of state change}
\label{weights_of_stat_change}
Besides the required duration ($\Delta$ T) to complete a state transition, there are also other parameters to weight a state transition, as well as an overall clinical pathway that consists of a set of state transitions. We use the duration, cost, comfort, and belief as weights to evaluate a state transition in the healthcare domain: 

\begin{itemize}
\item  Duration - a positive number which indicates how long the execution of a step takes.
\item  Cost - a positive number indicating how much the step costs in Euros.
\item Comfort - a number between 0 and 1 indicating how comfortable the step is for the patient, 1 being very comfortable, 0 being uncomfortable.
\item Belief - a number between 0 and 1 indicating the probability that the step actually leads to the expected result.
\end{itemize}

%\st{More details of weighted state transitions are %enclosed in our other paper ~\mbox{\cite{doerthe2021}} %This paper focuses on the application of weighted state %transition logic in the healthcare domain to predict %the future state, as well as applications built on top %of the predicted future state.}
The overall weights of a path is calculated by combing the weights of each action listed in the path. The overall weights of duration and cost are additive, by summing up the duration and cost of each action respectively. The overall weights of comfort and belief are multiplicative, by multiplying the comfort and belief of each actions respectively. The user can put constraints on overall weights of a target path during path generation, in order to limit the candidate path.

\subsection{Modeling the weighted state transition logic}

As already indicated above, our theory is based on a fragment of linear logic and extends the logic's concepts by the possibility of adding weights to the different transition rules. 

Given the disjoint countable sets $\mathcal{V}$ of variables, $\mathcal{C}$ of  constants, and $\mathcal{P}$ of predicates and the set $\mathcal{W}$ of weights, we define the language $\mathcal{L}$ of weighted state transition logic in Table~\ref{syntax}. 

\begin{table}

\begin{singlespace}
\begin{tabular}{llr}
\hline
Syntax: &&\\
&&\\
\texttt{t ::=}&&                    terms:\\
      & \texttt{v}\hspace{0.5\textwidth} &                variables\\
      & \texttt{c} &                constants\\
%      & \texttt{p} & predicates \\
     % & \texttt{e} &                 expressions\\
     % & \texttt{(k)}& lists\\
     % & \texttt{()}& empty list\\
      &&\\
\texttt{p}      && predicates\\
&&\\
\texttt{w}      && weights\\
&&\\
%\texttt{k ::=}&&                    listcontent\\  
%       &\texttt{t}  &\\
%       &\texttt{t k}&\\
%\texttt{e ::=}&&                    expressions:\\
%      &\texttt{<>} &                true\\
%       &\texttt{<f>} &               formula expression\\
%       &\texttt{<>} & true\\
%       &\texttt{false}       &               false\\
%       &&\\
\texttt{f ::= } & &                   basic formulas:\\  
    &  \texttt{p(t,t)}&                atomic formula\\
    
%    &  \texttt{f} $\rightarrow \bot$ &\\
%    &  \texttt{f} $\rightarrow \top$ &\\
%    &  $\top \rightarrow $ \texttt{f} &\\
%    &   $\bot \rightarrow $ \texttt{f} &\\
    & $ \texttt{f} \otimes \texttt{f}$ &    basic             conjunction\\
          &&\\
\texttt{e ::= } & &                   existential formulas:\\ 
   &  \texttt{f}     & simple formulas\\
    &  \(\exists\)\texttt{v.f}     & quantified formula\\
      &&\\
\texttt{g ::= } & &                   wtf formulas:\\ 
&  \texttt{e} & existential formula\\
&  (!\texttt{e}) $\multimap$ \texttt{f}& implication\\
    & $\texttt{(e} \otimes \texttt{(!e))}\multimap_{\texttt{w}} \texttt{f}$ & transition rule \\
     &  \(\forall\)\texttt{v.g}     & universal formula\\
      & $ \texttt{g} \otimes \texttt{g}$ &            conjunction\\
    \hline
\end{tabular}
\end{singlespace}
\caption{Syntax of the wtf language $\mathcal{L}$ over $\mathcal{V}\cup\mathcal{C}\cup\mathcal{P}\cup\mathcal{W}$.\label{syntax}}
\end{table}

If we ignore the weights at the transition rule (i.e. we only consider $\texttt{(e} \otimes \texttt{!e)}\multimap \texttt{f}$ instead of $\texttt{(e} \otimes \texttt{!e)}\multimap_{\texttt{w}} \texttt{f}$), the semantics of our logic is the same as for general linear logic~\cite{girard1987linear} with the only modification that the linear implication ($\multimap$) removes \emph{all} occurrences of the facts occurring in its antecedent (unless the antecedent or part of the antecedent stays with the exponential !) and not just one. The reason that we chose this modification is of practical nature:  if in a clinical set-up two sources provide the exact same fact, it is more likely that this fact is a duplicate (the patient will for example only have  one body temperature at one point in time) than that we actually have to  deal  with two additive values (the body temperature will not be the sum of two measured values as a linear implication would normally expect it).

Before we further discuss the meaning of the weights, we have a closer look at the restrictions we impose on our rules and on the alignment of our formal syntax with the concepts introduced earlier, especially with Figure~\ref{fig:state-transition}.

Our logic supports facts (atomic formulas) and conjunctions\footnote{Note that the linear conjunction $\otimes$ has the same meaning as the first order conjunction $\wedge$.} of facts (atomic conjunctions) just as they occur in first order logic. These facts can furthermore be quantified. 
%Here, it is important to note that we only allow universal quantification outside of existential quantification ($\forall x. \exists y.$) but not the opposite order ($\exists y.\forall x.$).
Our logic furthermore allows two kinds of rules: classical implications as we also find them in first order logic ((!\texttt{e}) $\multimap$ \texttt{f} in linear logic notation) and transition rules ($\texttt{(e} \otimes \texttt{(!e))}\multimap_{\texttt{w}} \texttt{f}$). The former kind of rules is used in our applications to model background knowledge (for example that a patient whose body temperature exceeds a certain limit has a fever), the latter kind of rules are the  transition descriptions informally introduced in Figure~\ref{fig:state-transition} above. Here, the consequence of  the  rule is used to indicate the To-state, the first conjunct  from the  antecedent represents the From-state, the second conjunct the condition which is still valid after the application of the rule (this is indicated by the linear exponential !). The weights are written  as an index of the implication. Our formal syntax does not include action names since these are not relevant from a logical perspective and are only present in our rules for practical reasons (when showing a care path to a user it is easier to refer to rules by names instead of listing the rules themselves).

Note, that both kinds of rules do not support existentially quantified consequences. This has mainly practical reasons, as existential rules can heavily impair the  performance of reasoning~\cite{existentials}.

As  a last point we further explain the  weights associated with each transition  rule. The weights \texttt{w} should be a vector of positive numbers with a fixed length $n$, that is $\texttt{w}\in\mathbb{R}_{\geq 0}^n$. In our implementation we use a vector of length 4, but this number can be changed depending on the use case. However, it is important for our logic that the  length of our weights vector is fixed and that the weights are stated for  each transition rule. Just as in classical linear logic, the transition rules can be applied to the set of facts and change the  current state. Additionally to these applications, we compose the overall weights of the resulting states. To do that, we need to define a start value $w_0\in \mathbb{R}_{\geq 0}^n$ and a transition function $t:\mathbb{R}_{\geq 0}^n\times \mathbb{R}_{\geq 0}^n\rightarrow \mathbb{R}_{\geq 0}^n$ which can be used to determine the weight for  each state after the application of a transition rule. If we have the weight $w_s$  of  a current state $s$ and we want to a apply a transition rule with the weight  $w_t$, we calculate the weight of the resulting state as $w_t=t(w_s,w_t)$. The weight of a new state always only depends on the weight of the previous state and the  weight of the transition rule. This property of our logic is very similar to the Markov property \cite{howard1960dynamic}.

In our concrete implementation as described above, the start value is defined as $w_0=(0,0,1,1)$, corresponding to the weights of duration, cost, comfort and belief defined in \ref{weights_of_stat_change}. Since the calculation of duration and cost is defined as additive, and the calculation of cost and comfort is defined as multiplicative, the transition function $t$ is defined as 
\[
t((x_1,x_2, x_3, x_4), (y_1, y_2, y_3, y_4))=(x_1+y_1, x_2+y_2, x_3\cdot y_3, x_4\cdot y_4)
\]
For each transition rule application, we calculate the weight of the new state by applying the function on the weight of the previous state and the weight of the transition rule.

\section{Implementation of weighted state transition logic - a semantic web based approach}

The weighted state transition logic presented in the previous section is implemented with semantic representation as backward rules in N3 language. We created an ontology named gps-schema \footnote{http://josd.github.io/eye/reasoning/gps/gps-schema} to enable a semantic representation of state changes. The current state of a patient is represented with RDF graphs, and the target to reach is represented as an N3 query. Background knowledge is also introduced as RDF graphs or N3 rules. We use the semantic reasoning engine EYE to execute a set of tasks such as path generation and path validation. To meet the special needs of those tasks, we created several plugins \footnote{https://github.com/hongsun502/wstLogic/tree/master/engine} and rule sets for that reasoner.

\subsection{Semantic description of state change} \label{semantic_description}

This section uses the neoadjuvant chemo-radiotherapy in the domain of colon cancer as an example to introduce the semantic description of state change. The introduction of background knowledge and target description also uses examples from the colon cancer treatment. The detailed examples of treating colon cancer can be found in our GitHub project \cite{wstLogic}.

Listing 1 shows the semantic description of a neoadjuvant chemo-radiotherapy in the domain of colon cancer. In general, this description indicates that by taking the action of neoadjuvant chemo-radiotherapy, the size of the tumor is expected to shrink to 70\% of its original size. 

Line 9 indicates the specialized domain of the action, in this case, it is care:Colon\_cancer. We use 'Map' to indicate domain information of each specific medical domain, mimicking a map that provides different paths. We use the concept 'Map' to separate different domain knowledge, so that domain experts can focus on creating rules in their own expertise. The graph stated in the From section (lines 11-12) indicates the state before the action is applied, it will be retracted once the action is started. In this case, the current tumor size and metastasis risk of the patient will then be retracted. Line 13 contains the transition state. It indicates that during the state transition, the patient is receiving neoadjuvant chemo-radiotherapy. The graph stated in the transition section will be asserted when the action is started. It will be retracted when the action of neoadjuvant chemo-radiotherapy is finished. Lines 14-15 indicate the target state. When the state transition is finished, new values of tumor size (?new\_size) and metastasis risk (?new\_risk) will be asserted. The new size and new risk are reflecting the expectation of the treatment. In reality, the new size and new risk might differ as the confidence of reaching the target is indicated by the parameter belief.

%\begin{minipage}{\linewidth}
\begin{lstlisting}[frame=bt,numbers=left,float=tp,basicstyle=\footnotesize,caption=Sample state change representation of colon cancer therapy]
PREFIX math: <http://www.w3.org/2000/10/swap/math#>
PREFIX xsd: <http://www.w3.org/2001/XMLSchema#>
PREFIX gps: <http://josd.github.io/eye/reasoning/gps/gps-schema#>
PREFIX action: <http://josd.github.io/eye/reasoning/gps/action#>
PREFIX sct: <http://snomed.info/id/>
PREFIX therapy: <http://josd.github.io/eye/reasoning/gps/therapy#>
PREFIX care: <http://josd.github.io/eye/reasoning/gps/care#>

{care:Colon_cancer                        #Map
  gps:description (
  {?patient care:tumor_size ?size.
   ?patient care:metastasis_risk ?risk.}  #From
  {?patient gps:therapy therapy:Neoadjuvant_chemoradiotherapy.} 
  {?patient care:tumor_size ?new_size. 
   ?patient care:metastasis_risk ?new_risk.}  #To
    action:Neoadjuvant_chemoradiotherapy  #Action
    "P50D"^^xsd:dayTimeDuration           #Duration
    14147                                 #Cost
    0.9                                   #Belief
    0.4                                   #Comfort
  )} <= 
{?patient care:diagnosis sct:363406005.   #Colon cancer  
 ?patient  care:tnm_t ?t_value .          #Tumor-Node-Metastasis
 ?t_value math:greaterThan 2 .        
 ?patient care:tumor_size ?size.
 (?size 0.7) math:product ?new_size.
 ?patient  care:metastasis_risk ?risk .
 (?risk 0.5) math:product ?new_risk. }.   #Condition graph
\end{lstlisting}
%\end{minipage}

Line 16 indicates the action to be taken in the state transition is neoadjuvant chemo-radiotherapy. Lines 17-20 indicate the weights of the state transition. Duration indicates the action will take 50 days. Cost indicates the cost of the action is 14147 Euros. It is believed 90\% chance the target can be reached, and the comfort level of this action is 40\%. Both Belief and Comfort are initially subjective values based on the inputs of physicians. Nevertheless, they can be based on existing studies, as well as being updated following the outcome of evaluating the actual outcome of the state transition. 

Lines 22-28 form the section of Condition. Lines 22-24 indicate the premise of carrying this action, that is a patient is diagnosed with colon cancer (line 22), and the tumor is reaching more than two layers of the colon (lines 23-24). The new size of tumor is calculated in lines 25-26, it will be 70\% of the original size. The new risk of metastasis is calculated in lines 27-28, it will be 50\% of the original risk. The current calculations of the target values are simplified for demonstration purpose. In clinical practice, the new metastasis risk can be calculated based on the detailed status of a patient, even including factors such as genetic variants or consulting external machine learning services.

\subsection{Semantic description of background knowledge}
Background knowledge, such as stating a drug conflict, or asserting a statement of fever when the body temperature is above 38 degrees, are expressed as N3 backward rules. Listing 2 shows a sample of semantic description of a conflict between the drug Pramipexol and surgery of colon cancer. 

\begin{lstlisting}[frame=bt,numbers=left,basicstyle=\footnotesize,caption=Sample semantic description of a conflict]
PREFIX gps: <http://josd.github.io/eye/reasoning/gps/gps-schema#>
PREFIX med: <http://josd.github.io/eye/reasoning/gps/medication#>
PREFIX surgery: <http://josd.github.io/eye/reasoning/gps/surgery#>

{ ?patient gps:alert 
  {medication:Pramipexol gps:conflict surgery:surgery_colon_cancer.}.
} <=
{ ?patient gps:medication med:Pramipexol.
  ?patient gps:surgery surgery:surgery_colon_cancer. }.
\end{lstlisting}

%\begin{figure*}
%\centering\includegraphics[width=1.0 \linewidth]{description.png}
%\caption{Step description}
%\label{fig:step-description}
%\end{figure*}
%
%
%
%\begin{figure*}
%\centering\includegraphics[width=0.95\linewidth]{goal.png}
%\caption{Goal description}
%\label{fig:goal-description}
%\end{figure*}

\subsection{Path generation}

\begin{figure*}[ht]
\centering\includegraphics[width=0.99\linewidth]{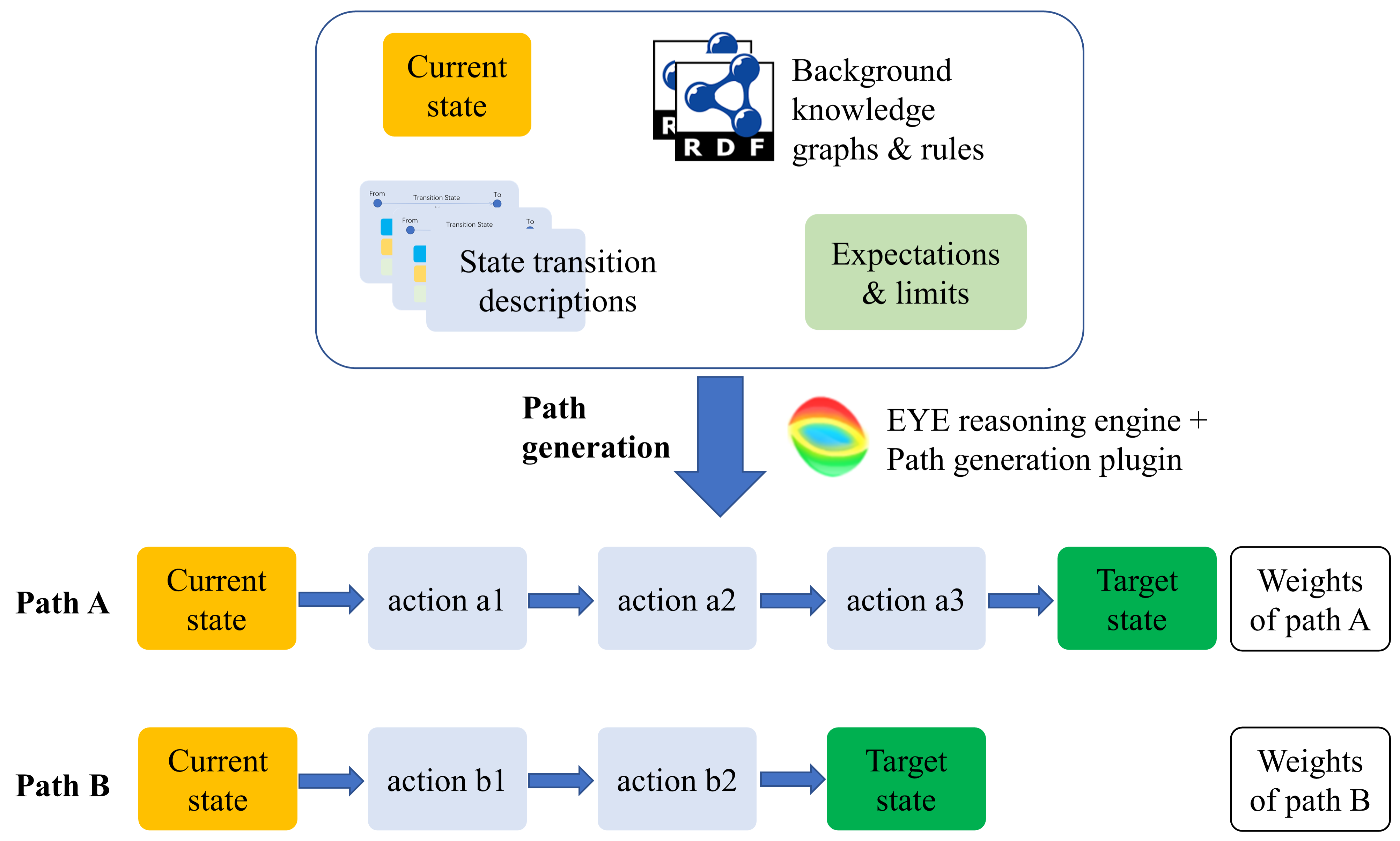}
\caption{Path generation}
\label{fig:path-generation}
\end{figure*}

Once the state transitions and background knowledge of a relevant domain are set up, it is possible to automatically generate a set of potential paths from the current state of a patient towards a target state. Figure \ref{fig:path-generation} shows the process of path generation. The EYE reasoning engine takes the current state of the patient, together with background knowledge and state transition descriptions of the relevant domain as inputs of a reasoning process for path generation. The expectations and constraints of a target path are expressed as the query of the process. The path generation plugin \footnote{https://github.com/hongsun502/wstLogic/tree/master/engine/gps-plugin.n3} finds a set of possible paths that start from the current state and end with the target states, with the stated constraints in the query. The path search process is carried out as forward chaining.

\subsubsection{Semantic description of a target}
Listing 3 shows an example of the target description of a colon cancer therapy. Lines 7-10 describe the target: the tumor size of the patient should be 0, and the metastasis risk should be lower than 10\%. Line 11 defines the actions of a path (?PATH), as well as the overall duration, cost, belief, and comfort to be calculated for a path. Line 12 defines the constraints of a path, the maximum duration (150 days), maximum cost (50000 Euros), the minimum overall belief (0.1), and the minimum overall comfort (0.1). Line 14 passes the generated paths to the output. It consists of the action sets, duration, cost, belief, comfort, as well as the metastasis risk by the end of the path.

\begin{lstlisting}[frame=bt,numbers=left,float=tp,basicstyle=\footnotesize,caption=Sample target description of a colon cancer therapy]
PREFIX math: <http://www.w3.org/2000/10/swap/math#>
PREFIX xsd: <http://www.w3.org/2001/XMLSchema#>
PREFIX gps: <http://josd.github.io/eye/reasoning/gps/gps-schema#>
PREFIX care: <http://josd.github.io/eye/reasoning/gps/care#>

{?SCOPE gps:findpath (
  { ?patient a  care:Patient.  
    ?patient  care:tumor_size 0 . 
    ?patient  care:metastasis_risk ?risk .
    ?risk math:lessThan 0.1 . }        
    ?PATH ?DURATION ?COST ?BELIEF ?COMFORT 
    ("P150D"^^xsd:dayTimeDuration 50000.0 0.1 0.1)).
}=> {
 ?patient gps:path (?PATH ?DURATION ?COST ?BELIEF ?COMFORT (?risk)).}
\end{lstlisting}

\subsubsection{Sample paths}

\begin{figure*}[ht]
\centering\includegraphics[width=0.99\linewidth]{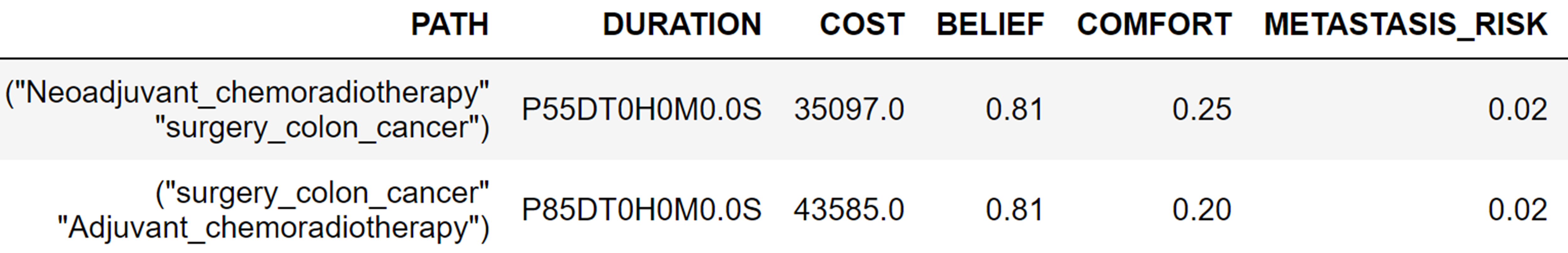}
\caption{Sample paths}
\label{fig:paths}
\end{figure*}

%\begin{figure}[h]
%\centering
%\caption{Sample paths}
%
% \begin{tabular}{c  c  c  c  c  c} 
% \hline
% PATH & DURATION & COST & BELIEF & COMFORT & METASTASIS_RISK \\
% \hline
% ("Neoadjuvant_chemoradiotherapy" "surgery_colon_cancer") & %P55DT0H0M0.0S & 35097.0 & 0.81  & 0.25  & 0.02 \\ 
% \hline
% ("surgery_colon_cancer" "Adjuvant_chemoradiotherapy") & P85DT0H0M0.0S %& 43585.0 & 0.81 & 0.20 & 0.02 \\
% \hline
% \end{tabular}
% \label{figure:paths}
%\end{figure}

Figure \ref{fig:paths} shows two sample paths generated in the path generation, corresponding to the target stated in Listing 3. The column 'PATH' indicates the actions to be taken in the clinical pathway. The 'METASTASIS\_RISK' indicates the risk of metastasis. The rest of the columns are overall weights of a path. The overall duration and cost are calculated by summing up the duration and cost of each action listed in the path. The overall belief and comfort are calculated by multiplying the belief and comfort of each action. The path 0 first takes neoadjuvant chemo-radiotherapy, followed by surgery of colon cancer. The path 1 first takes surgery of colon cancer, followed by adjuvant chemo-radiotherapy. It can be observed that the first path has a shorter duration, lower cost and better comfort.

\subsection{Path validation}

\begin{figure}[ht]
\centering\includegraphics[width=0.99\linewidth]{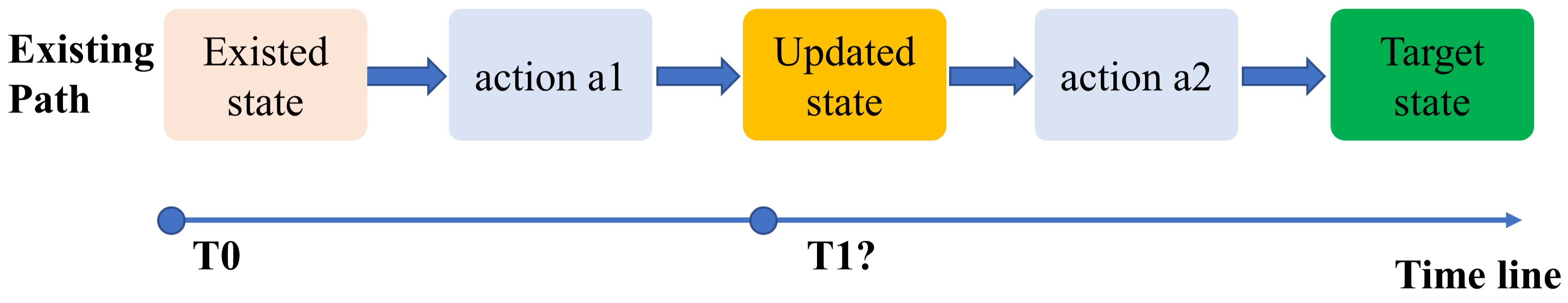}
\caption{Path validation}
\label{fig:path-validation}
\end{figure}

Path validation checks whether executing a planned path would lead to the defined goal following the update of a patient state. Figure \ref{fig:path-validation} shows that a path is generated at T0, with the known patient state at T0. After the first planned action is carried out, the patient state is updated at T1. Path validation is executed at T1 to check if the target state can still be reached.

The path validation takes the up to date patient state and performs the state transitions of the planned path sequentially till the end of the path. By applying the state transitions, it predicts the future state and checks whether the goal can be fulfilled or not. In the example given in Figure \ref{fig:path-validation}, the path validation process takes the updated state (at T1), and applies the state transition of action a2 to generate a predicted state, and check if the predicted state fulfils the target state. In case a target state is predicted as no longer reachable, the responsible physician will get notified before action a2 actually takes place.

\subsection{Conflict detection}

\begin{figure}[ht]
\centering\includegraphics[width=0.99\linewidth]{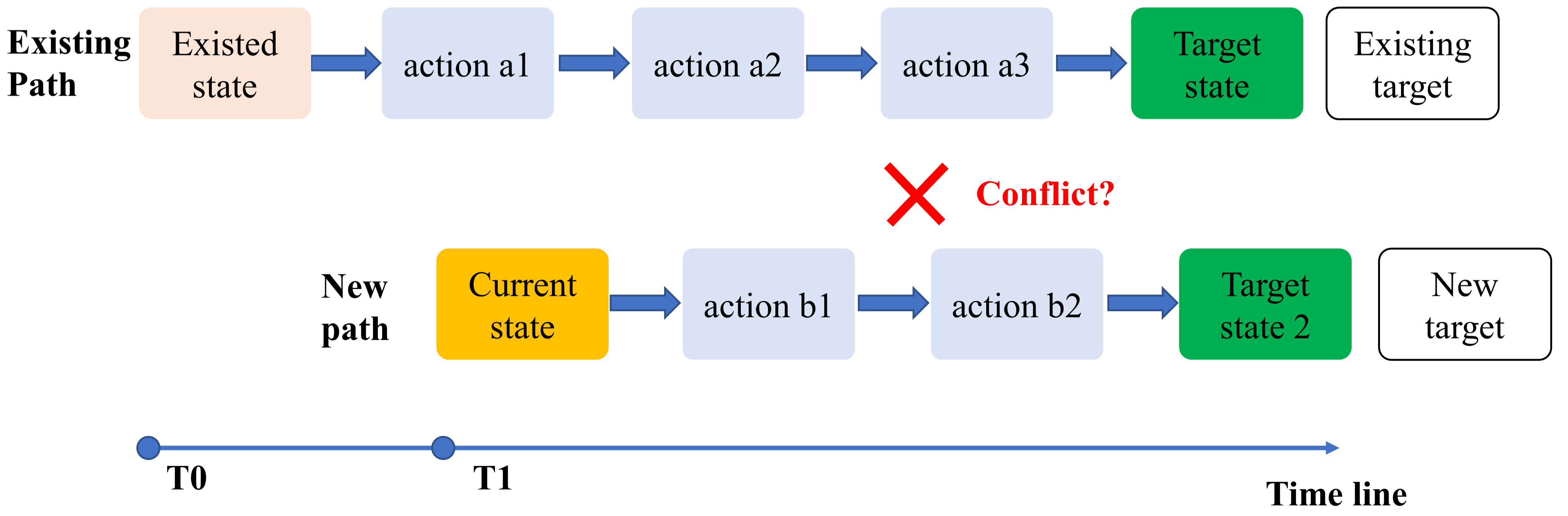}
\caption{Conflict detection}
\label{fig:conflict-detection}
\end{figure}

Conflict detection checks if a new path brings any conflict with existing paths. Figure \ref{fig:conflict-detection} shows that at T1, a new path is generated. The original sequence of the existing path is (a1,a2,a3). With the extension of the new path, the sequence of the aggregated path would be (a1,a2,b1,a3,b2). Similar to the path validation, conflict detection performs the state transitions of the aggregated path sequentially to predict the future states and search for conflicts. It checks two types of conflicts: firstly, it checks if there are any explicit conflicts between different operations, e.g., the conflict stated in Listing 2. 

Secondly, it checks implicit conflicts. Both existing path and new path are predicted to be able to reach their targets when they are executed respectively. There is no guarantee that the goals can still be reached with the interference from the other path. Implicit conflict detection checks if both the existing target and the new target can still be reached by applying the aggregated path sequentially.

Both path validation and conflict detection aim to predict if there is a potential failure of a clinical pathway in the future, based on the future states build on state transitions. It allows the early intervention of the clinical pathway before irreversible actions takes place.

\section{GPS for integrated care with weighted state transition logic}
The methods introduced in the previous section were implemented in the GPS4IC (GPS4IntegratedCare) project \cite{GPS4IC}. The GPS4IC project aims to develop a platform which allows the automatic generation of dynamic and personalized clinical pathways. A smart workflow engine was developed in this project which is able to dynamically generate personalized clinical pathways based on weighted state transition logic. The engine is also able to aggregate different clinical pathways, detect conflicts, and validate ongoing pathways.

\subsection{Architecture}

%Figure \ref{fig:workflow} shows the limitations of the current state %of the art and how the adaptive pathways with linear logic addresses %these challenges.
%
%
%
%\begin{figure*}
%\centering\includegraphics[width=0.75\linewidth]{workflow.png}
%\caption{Current fixed pathways vs. proposed dynamic pathways with %linear logic}
%\label{fig:workflow}
%\end{figure*}

\begin{figure*}[ht]
\centering\includegraphics[width=0.95\linewidth]{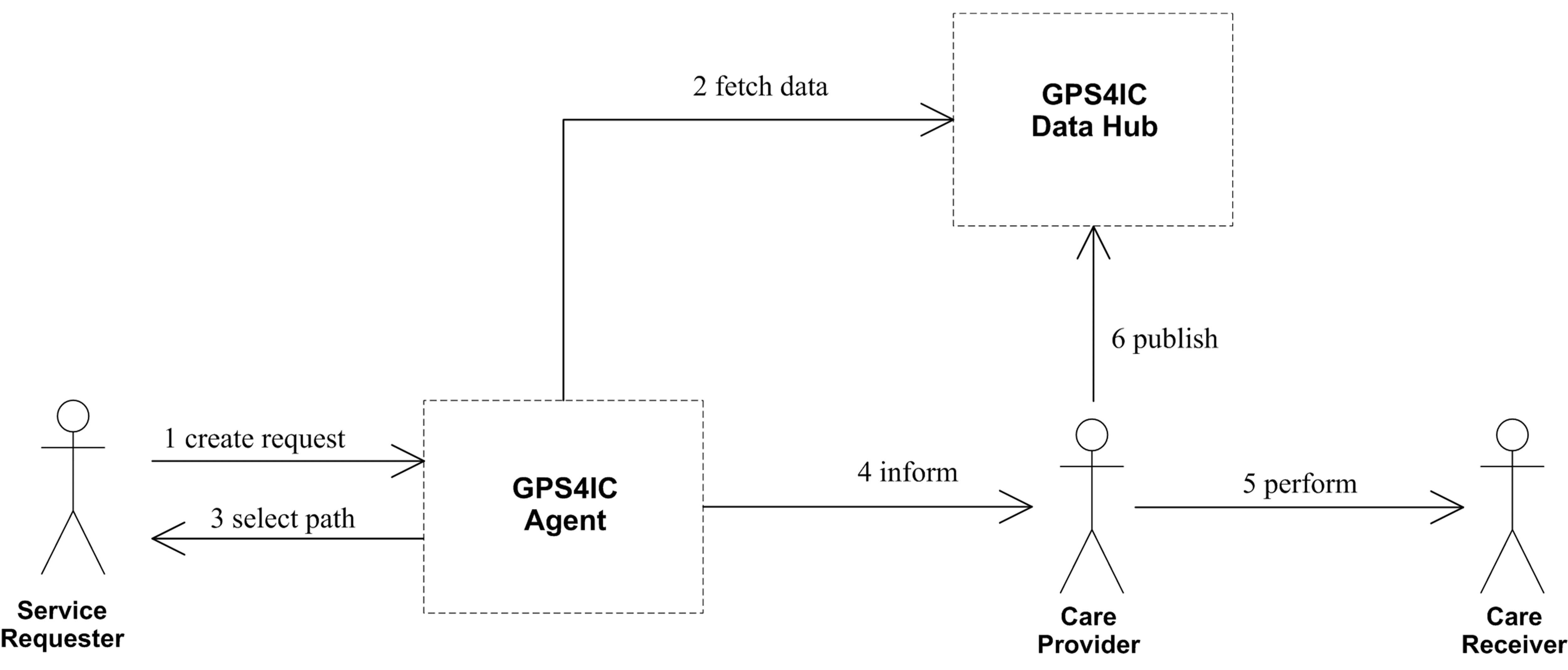}
\caption{Solution architecture of GPS4IC project}
\label{fig:black-box}
\end{figure*}

Figure \ref{fig:black-box} shows the solution architecture of the GPS4IC project. The GPS4IC platform consists of the GPS4IC agent and the GPS4IC data hub. 

The GPS4IC data hub provides relevant data to the GPS4IC agent. The data hub retrieves live data from different sources to reflect the latest state (of a patient). We use the semantic data virtualization approach \cite{sun2015semantic} to build the data hub for the following benefits:
\begin{itemize}
  \item The data is still kept in their original repositories, which is crucial in clinical applications where security and privacy are important.
  \item The data is semantically processed, so that it can be used together with the state change descriptions by the semantic reasoning engine.
\end{itemize}

Besides providing data to reflect the state, the data hub also provides relevant knowledge including:
\begin{itemize}
  \item State transition descriptions, which are fundamental elements to process a clinical pathway.
  \item Clinical knowledge, e.g., statements of contraindications or hierarchical relationships with clinical concepts, which are used to infer further knowledge or detect conflicting situations.
\end{itemize}

The GPS4IC agent communicates with the service requester to specify goals, and confirms the path to take. It retrieves data from the GPS4IC data hub to run the path generation, and it communicates the requested action to the care provider. The GPS4IC agent is able to accomplish the use cases listed in the following section. Once a care provider finishes a task, the relevant update is made in corresponding local systems, and is reflected back to the GPS4IC agent through the data hub. For example, if a temperature measurement is made by a nurse, the value of the temperature will be entered into the EHR system, and when the GPS4IC agent requests the latest temperature, the data hub will provide this newly entered value.

There are different types of care providers involved in providing care to the care receiver. Their interactions with the GPS4IC agent are therefore also different. For the executioners of a planned action, such as the nurse or the surgeon, they could get a notification from the GPS4IC agent when a planned action is to be executed. A treating physician whom is responsible for the care-path of a patient interacts with the GPS4IC agent more frequent. First, when a new path is needed, the GPS4IC agent returns a set of candidate paths to the treating physician to choose from. Secondly, when the treating physician selects one path, the GPS4IC agent checks if there are any conflicts with any of the existing paths, and returns the outcome to the treating physician. Lastly, when the state of a patient is updated, the GPS4IC agent checks if all the existing paths can still reach their targets. If any of the existing paths fails to reach its target after the state update, the corresponding treating physician will get informed by the GPS4IC agent.

\subsection{Scenario - Pathway management of Parkinson and Colon Cancer }
The scenario of managing personalized clinical pathways for a patient with comorbidity of colon cancer and Parkinson disease has been implemented in the GPS4IC project. The scenario represents a simplified use case of chemotherapy and surgery treatment of colon cancer, in combination with the comorbidity of Parkinson disease. The sequence of applying chemotherapy and performing surgery in colon cancer treatment is influenced by whether the colon is blocked by the tumor, such a constraint is expressed explicitly in the condition section of the state change description. The sequence in colon cancer treatment, in combination with the TNM (tumor-node-metastasis) status, largely influences the five year survival rate and post surgery relapse risk of colon cancer treatment. Those influences are expressed as background knowledge with a set of N3 rules. Similarly, the calculation of Unified Parkinson's Disease Rating Scale (UPDRS) in Parkinson treatment is also expressed as N3 rules. The scenario is first described by two medical doctors in the GPS4IC project, then reviewed by the clinical advisory board of the project, and finally implemented with the weighted state transition logic. It is published as an open source project on GitHub \cite{wstLogic}. The corresponding data and domain knowledge, represented with N3 language, are enclosed in the GitHub project. Source code of the path generation, path validation, and conflict detection engines are also provided. The scenario of path generation and validation of each disease, and conflict detection between two paths are also constructed with visualizations. The scenario demonstrates the weighted state transition logic presented in this paper is capable of taking into account the constraints of clinical procedures to generate a personalized care-path. It also shows the potential of such an approach to integrate with existing clinical knowledge with semantic web technology.

\subsection{Performance measurement}
The weighted state transition logic presented in this paper is designed to carry the tasks of path generation, path validation, and conflict detection. Path validation checks whether a given path can reach the target state by sequentially applying the state changes according to the actions listed in the path. The complexity of such calculations is linear to the number of actions listed in the path. The conflict detection first merges the actions listed in different care-paths into an aggregated path, and then checks whether there is any conflict, as well as whether the aggregated path can still lead to each of the targeted states of the original paths. The complexity of such calculations is also linear to the number of actions listed in the aggregated path. The path validation and conflict detection performed in the scenario are both executed in less than 10 ms where 2 diseases and 3 actions are listed in the aggregated path. Since the complexity is linear, it is expected that the weighted state transition logic is capable to execute path validation and conflict detection for complex situations, such as 5 comorbidities with each 6 actions, in around 100 ms.

The process of path generation requires more complex computation to explore different possible paths that lead from the current state to the target state. Nevertheless, since the care-path is generated for each disease independently and aggregated afterwards, the impact of generating care-paths for multiple diseases is therefore still linear. The complexity of path generation is largely influenced by the number of state change descriptions (of candidate actions). As introduced in section 4.1, we use 'Map' to group state change descriptions in each specific medical domain, which largely reduce the number of potential state changes in path generation. A benchmark is made to measure the performance of path generation with a different numbers of candidate actions. We automatically generate a set of state change descriptions with 'From' and 'To' both randomly assigned as one of 10 candidate states. We apply path generation to query possible paths from the initial state to a target state. Details of the benchmark design are available at: \emph{https://github.com/hongsun502/wstLogic/tree/master/benchmark}. 

\begin{table}[h!]
\centering
\caption{Benchmark of path generation (average times of 10 runs)}

 \begin{tabular}{c | c | c} 
 \hline
 number of candidate actions & number of paths & inference time (ms) \\
 \hline\hline
 10  & 0  & 0.3 \\ 
 \hline
 100  & 1  & 3 \\
 \hline
 1000  & 15  & 411 \\
 \hline
 10000  & 29  & 42003 \\
 \hline
 \end{tabular}
 \label{table_benchmark}
\end{table}

The benchmark is executed on a laptop with a CPU of AMD Ryzen 5 3550H processor @ 2.1 GHz, and the results are summarized in Table \ref{table_benchmark}. Although the performance of path generation is nonlinear, we consider in a real-world application the candidate actions in a disease domain are limited and the weighted state change logic is capable of generating paths in a timely manner. The path generation for the colon cancer and Parkinson disease in the scenario presented in Section 5.2 are both within 10 ms range. In addition, when the performance is jeopardized with too many possible paths, it is possible to limit the number of possible paths by adjusting the path limit constraint and thus improve the path generation speed.

\subsection{Lessons learned}

During the GPS4IC project, several workshops were held in different hospitals to demonstrate the use cases to physicians and nurses for feedback. In order to cope with comorbidity, we allow the users of our system to choose different care plans per disease, and validate if there are conflicts between different pathways when a new pathway is added or the patient state is updated. Initially, we planned in the project to automatically compose joint care plans for all diseases of a patient by stating a combined target to cope with multiple diseases. However, the user studies of the project showed that physicians deciding over a care plan prefer to only take responsibility for the diseases falling under their expertise. It was therefore decided that for each disease an independent care plan will be generated. This decision made it necessary to check different care plans for compatibility after a new clinical pathway is planned. If two care plans are compatible, they can be executed in parallel. If plans are not compatible, the persons responsible for the care plans can contact each other to come to a common agreement. Instead of frequently holding joint meetings to discuss treatment of comorbity in the current practice, our implementation can ease the communication between doctors and make them aware of each other's plans.

During the demonstrations in hospitals, it turned out that nurses are more focused on operational details, e.g., making an appointment for a planned surgery, while the physicians would like to avoid those operational details and focus on abstracted workflows. They found the inference based on background knowledge useful, and they also consider such a tool beneficial to bring the whole picture of ongoing clinical pathways to general practitioners.

It also demonstrated that there is a different acceptance level of such a tool in different disease domains. Neurologists found it is seldom that the consequence of a treatment can be explicitly stated as state change described in this paper. Radiologists share the concern of the difficulty in defining state change descriptions. However, they tend to be more tolerant to consider manually created state change descriptions, and let the description evolve following the feedback received on this system is a feasible approach. We therefore consider that the application of this tool should be limited to domains where explicitly state change descriptions are possible. 

\subsection{Limitations}

We implemented the scenario of pathway management with weighted state transition logic on a simplified clinical use case, and the approach is ready to be validated in real world clinical settings. However, such a real world validation is not yet done in the GPS4IC project yet. The main challenge is to extend the state change descriptions to include more actions in relevant domains. Nevertheless, following the lessons learned from the GPS4IC project, it is feasible to create the relevant state change descriptions.

The process of path generation is a Markov decision process that at each step, the path generation engine may take any of the executable actions, which leads to non-linear behaviour. We solved this issue by introducing the concept of 'Map' to limit the search scope in the path generation process. We also limit the executable actions by imposing more strict constraints in the condition section. By specifying more strict limit constraints on the overall weights in the target also helps to reduce the search space and speed up the path generation process. Lastly, by preselecting state change descriptions that are relevant to the target with goal based backward reasoning, we can largely reduce the amount of state change descriptions in path generation and therefore improve the speed.

\section{Conclusion}

This paper introduced the weighted state transition logic and its application in predicting future states for adaptive clinical pathway management. The traditional clinical pathways suffer from either lacking the flexibility to adapt, or lacking a pathway relying on the outcome of each step. Along with the increased population of elderly people, the number of patients with comorbidity will also increase dramatically. Traditional clinical pathways are invented to provide generalized guidelines to a specific disease, but they are not designed to provide personalized care that copes with comorbidity. Applying the state change concept of weighted state transition logic into the clinical domain, it allows us to model the consequences of each step in a clinical pathway, which eventually allows us to predict the future state in defining a care plan. This predicted future state is not based on a single clinical pathway, but takes into account the consequences of the planned steps on all the existing pathways. Building the system on top of a semantic web language (N3) also allows easy integration of existing knowledge and enables the generation of the personalized clinical pathway. 

The proposed approach is implemented in the GPS4IC project. The platform built in the project is able to generate a personalized clinical pathway, detect conflicts that are predicted to happen in the future, and validate ongoing clinical pathways. Future work will focus on improving the system following the outcomes learned from the GPS4IC project, i.e., to investigate better ways of creating state change descriptions, as well as fine-tuning the state change descriptions following the feedback received when the system is under operation. 

\section{Acknowledgement}
The described research activities were funded by Ghent University, imec, Flanders Innovation \& Entrepreneurship (VLAIO), and the European Union. More specifically, our research findings were established through the GPS4IC project \cite{GPS4IC}. Furthermore, the authors would like to thank Dirk Colaert for initiating the GPS4IC project and the contributions from Els Lion, Elric Verbruggen, Joachim Van Herwegen, Ruben Verborgh, and Estefania Serral during the project.

\section*{References}

\bibliography{ref}

\begin{thebibliography}{10}
\expandafter\ifx\csname url\endcsname\relax
  \def\url#1{\texttt{#1}}\fi
\expandafter\ifx\csname urlprefix\endcsname\relax\def\urlprefix{URL }\fi
\expandafter\ifx\csname href\endcsname\relax
  \def\href#1#2{#2} \def\path#1{#1}\fi

\bibitem{kinsman2010clinical}
L.~Kinsman, T.~Rotter, E.~James, P.~Snow, J.~Willis, What is a clinical
  pathway? development of a definition to inform the debate, BMC medicine 8~(1)
  (2010) 1--3.

\bibitem{de2006defining}
L.~De~Bleser, R.~Depreitere, K.~D. Waele, K.~Vanhaecht, J.~Vlayen, W.~Sermeus,
  Defining pathways, Journal of nursing management 14~(7) (2006) 553--563.

\bibitem{rotter2010clinical}
T.~Rotter, L.~Kinsman, E.~L. James, A.~Machotta, H.~Gothe, J.~Willis, P.~Snow,
  J.~Kugler, Clinical pathways: effects on professional practice, patient
  outcomes, length of stay and hospital costs, Cochrane database of systematic
  reviews~(3).

\bibitem{deneckere2012care}
S.~Deneckere, M.~Euwema, P.~Van~Herck, C.~Lodewijckx, M.~Panella, W.~Sermeus,
  K.~Vanhaecht, Care pathways lead to better teamwork: results of a systematic
  review, Social science \& medicine 75~(2) (2012) 264--268.

\bibitem{rotter2019clinical}
T.~Rotter, R.~B. de~Jong, S.~E. Lacko, U.~Ronellenfitsch, L.~Kinsman, Clinical
  pathways as a quality strategy, Improving healthcare quality in Europe.

\bibitem{chevalley2002osteoporosis}
T.~Chevalley, P.~Hoffmeyer, J.-P. Bonjour, R.~Rizzoli, An osteoporosis clinical
  pathway for the medical management of patients with low-trauma fracture,
  Osteoporosis international 13~(6) (2002) 450--455.

\bibitem{takegami2003impact}
K.~Takegami, Y.~Kawaguchi, H.~Nakayama, Y.~Kubota, H.~Nagawa, Impact of a
  clinical pathway and standardization of treatment for acute appendicitis,
  Surgery today 33~(5) (2003) 336--341.

\bibitem{berners2008n3logic}
T.~Berners-Lee, D.~Connolly, L.~Kagal, Y.~Scharf, J.~Hendler, N3logic: A
  logical framework for the world wide web, Theory and Practice of Logic
  Programming 8~(3) (2008) 249.

\bibitem{eye}
{Euler Yet another proof Engine - EYE }, \url{https://github.com/josd/eye},
  [Online; accessed January-2021].

\bibitem{GPS4IC}
{GPS4IntegratedCare Project Leaflet},
  \url{https://drupal.imec-int.com/sites/default/files/inline-files/imec.icon_leaflet_GPS4integratedcare_FINAL_WEB.pdf},
  [Online; accessed January-2021].

\bibitem{sun2009promises}
H.~Sun, V.~De~Florio, N.~Gui, C.~Blondia, Promises and challenges of ambient
  assisted living systems, in: 2009 Sixth International Conference on
  Information Technology: New Generations, IEEE, 2009, pp. 1201--1207.

\bibitem{campbell1998integrated}
H.~Campbell, R.~Hotchkiss, N.~Bradshaw, M.~Porteous, Integrated care pathways.,
  BMJ: British Medical Journal 316~(7125) (1998) 133.

\bibitem{HUANG2016227}
Z.~Huang, W.~Dong, L.~Ji, C.~He, H.~Duan, Incorporating comorbidities into
  latent treatment pattern mining for clinical pathways, Journal of Biomedical
  Informatics 59 (2016) 227--239.

\bibitem{colaert2007}
D.~Colaert, {Bringing the pieces together}, in: Towards Semantic
  Interoperability in e-Health Workshop, 2007.

\bibitem{chen2004towards}
H.~Chen, D.~Colaert, J.~De~Roo, A.~Healthcare, Towards adaptable clinical
  pathway using semantic web technology, in: W3C Workshop Semantic Web for Life
  Science, 2004.

\bibitem{sun2015semantic}
H.~Sun, K.~Depraetere, J.~De~Roo, G.~Mels, B.~De~Vloed, M.~Twagirumukiza,
  D.~Colaert, Semantic processing of ehr data for clinical research, Journal of
  biomedical informatics 58 (2015) 247--259.

\bibitem{zhang2016integrating}
Y.-F. Zhang, Y.~Tian, T.-S. Zhou, K.~Araki, J.-S. Li, Integrating hl7 rim and
  ontology for unified knowledge and data representation in clinical decision
  support systems, Computer methods and programs in biomedicine 123 (2016)
  94--108.

\bibitem{alexandrou2010holistic}
D.~Alexandrou, I.~Skitsas, G.~Mentzas, A holistic environment for the design
  and execution of self-adaptive clinical pathways, IEEE Transactions on
  Information Technology in Biomedicine 15~(1) (2010) 108--118.

\bibitem{swrl}
{SWRL}, \url{https://www.w3.org/Submission/SWRL/}, [Online; accessed
  January-2021].

\bibitem{wang2013creating}
H.-Q. Wang, J.-S. Li, Y.-F. Zhang, M.~Suzuki, K.~Araki, Creating personalised
  clinical pathways by semantic interoperability with electronic health
  records, Artificial intelligence in medicine 58~(2) (2013) 81--89.

\bibitem{rajkomar2018scalable}
A.~Rajkomar, E.~Oren, K.~Chen, A.~M. Dai, N.~Hajaj, M.~Hardt, P.~J. Liu,
  X.~Liu, J.~Marcus, M.~Sun, et~al., Scalable and accurate deep learning with
  electronic health records, NPJ Digital Medicine 1~(1) (2018) 18.

\bibitem{cutillo2020machine}
C.~M. Cutillo, K.~R. Sharma, L.~Foschini, S.~Kundu, M.~Mackintosh, K.~D. Mandl,
  Machine intelligence in healthcare—perspectives on trustworthiness,
  explainability, usability, and transparency, NPJ Digital Medicine 3~(1)
  (2020) 1--5.

\bibitem{bradbrook2005ai}
K.~Bradbrook, G.~Winstanley, D.~Glasspool, J.~Fox, R.~Griffiths, Ai planning
  technology as a component of computerised clinical practice guidelines, in:
  Conference on Artificial Intelligence in Medicine in Europe, Springer, 2005,
  pp. 171--180.

\bibitem{mcdermott1998pddl}
D.~McDermott, M.~Ghallab, A.~Howe, C.~Knoblock, A.~Ram, M.~Veloso, D.~Weld,
  D.~Wilkins, Pddl-the planning domain definition language (1998).

\bibitem{sutton2003syntax}
D.~R. Sutton, J.~Fox, The syntax and semantics of the pro forma guideline
  modeling language, Journal of the American Medical Informatics Association
  10~(5) (2003) 433--443.

\bibitem{alaboud2019personalized}
F.~K. Alaboud, A.~Coles, Personalized medication and activity planning in
  pddl+, in: Proceedings of the International Conference on Automated Planning
  and Scheduling, Vol.~29, 2019, pp. 492--500.

\bibitem{fox2002pddl+}
M.~Fox, D.~Long, Pddl+: Modeling continuous time dependent effects, in:
  Proceedings of the 3rd International NASA Workshop on Planning and Scheduling
  for Space, Vol.~4, 2002, p.~34.

\bibitem{verborgh2017pragmatic}
R.~Verborgh, D.~Arndt, S.~Van~Hoecke, J.~De~Roo, G.~Mels, T.~Steiner,
  J.~Gabarro, The pragmatic proof: Hypermedia api composition and execution,
  Theory and Practice of Logic Programming 17~(1) (2017) 1--48.

\bibitem{verborgh2014serendipitous}
R.~Verborgh, Serendipitous web applications through semantic hypermedia, Ph.D.
  thesis, Ghent University (2014).

\bibitem{gurvits2005markov}
L.~Gurvits, J.~Ledoux, Markov property for a function of a markov chain: A
  linear algebra approach, Linear algebra and its applications 404 (2005)
  85--117.

\bibitem{howard1960dynamic}
R.~A. Howard, Dynamic programming and markov processes.

\bibitem{van1986intuitionistic}
D.~Van~Dalen, Intuitionistic logic, in: Handbook of philosophical logic,
  Springer, 1986, pp. 225--339.

\bibitem{girard1987linear}
J.-Y. Girard, Linear logic, Theoretical computer science 50~(1) (1987) 1--101.

\bibitem{girard1995linear}
J.-Y. Girard, Linear logic: its syntax and semantics, in: Proceedings of the
  workshop on Advances in linear logic, 1995, pp. 1--42.

\bibitem{combi2014representing}
C.~Combi, M.~Gambini, S.~Migliorini, R.~Posenato, Representing business
  processes through a temporal data-centric workflow modeling language: An
  application to the management of clinical pathways, IEEE Transactions on
  Systems, Man, and Cybernetics: Systems 44~(9) (2014) 1182--1203.

\bibitem{existentials}
J.-F. Baget, M.-L. Mugnier, S.~Rudolph, M.~Thomazo, Walking the complexity
  lines for generalized guarded existential rules, in: Twenty-Second
  International Joint Conference on Artificial Intelligence, 2011.

\bibitem{wstLogic}
{Weighted State Transition Logic - wstLogic },
  \url{https://github.com/hongsun502/wstLogic}, [Online; accessed
  January-2021].

\end{thebibliography}
%%\bibliography{reference}

\end{document}